\documentclass[nofootinbib,twocolumn]{revtex4}
\input psfig.sty
\usepackage[dvips]{color}
\usepackage{graphicx}
\usepackage{epsfig}

\begin{document}

\title{Observational tests of non-adiabatic Chaplygin gas}
\author{S. Carneiro$^{1}$\footnote{saulo.carneiro@pq.cnpq.br} and C. Pigozzo$^{1,2}$\footnote{cpigozzo@ufba.br}}

\affiliation{$^{1}$Instituto de F\'{\i}sica, Universidade Federal da Bahia, Campus de Ondina, Salvador, BA, 40210-340, Brazil\\ $^{2}$Imperial Centre for Inference and Cosmology, Imperial College, Blackett Laboratory, Prince Consort Road, London SW7 2AZ, UK}


\begin{abstract}
In a previous paper \cite{CB} it was shown that any dark sector model can be mapped into a non-adiabatic fluid formed by two interacting components, one with zero pressure and the other with equation-of-state parameter $\omega = -1$. It was also shown that the latter does not cluster and, hence, the former is identified as the observed clustering matter. This guarantees that the dark matter power spectrum does not suffer from oscillations or instabilities. It applies in particular to the generalised Chaplygin gas, which was shown to be equivalent to interacting models at both background and perturbation levels. In the present paper we test the non-adiabatic Chaplygin gas against the Hubble diagram of type Ia supernovae, the position of the first acoustic peak in the anisotropy spectrum of the cosmic microwave background and the linear power spectrum of large scale structures. We consider two different samples of SNe Ia, namely the Constitution and SDSS compilations, both calibrated with the MLCS2k2 fitter, and for the power spectrum we use the 2dFGRS catalogue. The model parameters to be adjusted are the present Hubble parameter, the present matter density and the Chaplygin gas parameter $\alpha$. The joint analysis best fit gives $\alpha \approx - 0.5$, which corresponds to a constant-rate energy flux from dark energy to dark matter, with the dark energy density decaying linearly with the Hubble parameter. The $\Lambda$CDM model, equivalent to $\alpha = 0$, stands outside the $3\sigma$ confidence interval. This result is still valid if we use, as supernovae samples, the SDSS and Union2.1 compilations calibrated with the SALT2 fitter.
\end{abstract}

\maketitle

\section{Introduction}
\label{Introduction}

In spite of the amount of precise cosmological data and good concordance between different observations, understanding the two dominant cosmic components, dark matter and dark energy, is still in order. There are suggestions of unifying the dark sector, and a known example is the generalised Chaplygin gas model, in which the dark sector is described by a unique component with density $\rho$ and negative pressure 
\begin{equation} \label{EoS}
p = -A/\rho^{\alpha},
\end{equation}
where $A > 0$ and $\alpha > -1$ are constants \cite{GCG1}. In early times it behaves like pressureless matter, and tends to a cosmological constant in the asymptotic future. However, the presence of unobserved oscillations and instabilities in the gas power spectrum was soon realised \cite{Sandvik}.

The possibility of unified descriptions of the dark sector is related to the so called dark degeneracy, i.e. the fact that the dark components are not uniquely defined \cite{Kunz}. Nevertheless, in a recent paper \cite{CB} it was argued that in observational tests it is necessary to properly define the clustering component of the dark sector, the component observed in large scale structures. Any dark fluid, unified or not, can be mapped into two interacting components, the first with zero pressure (here named dark matter) and the other with equation of state $p_{\Lambda} = -\rho_{\Lambda}$ (here called dark energy). It was shown that, if the former is non-relativistic, the latter does not cluster and, hence, the defined dark matter can be identified with clustering matter. In this case, the dark matter power spectrum does not suffer from oscillations or instabilities, which are avoided owing to the presence of late-time entropic perturbations related to the energy flux between the components.

These results apply in particular to the generalised Chaplygin gas (GCG). Its split into two interacting components permits the identification of its clustering part and hence a proper analysis of observations \cite{CB,GCG,Wands}. The clustering component has no oscillations or instabilities in its power spectrum. The GCG parameter $\alpha$ determines the rate of energy flux between the components, with a negative $\alpha$ meaning creation of dark matter at the expenses of dark energy.

In this paper we perform an observational analysis of the non-adiabatic GCG, by testing it against precise cosmological observations, namely the Hubble diagram of type Ia supernovas (SNe Ia), the position of the first acoustic peak in the anisotropy spectrum of the cosmic microwave background (CMB) and the distribution of large scale structures (LSS). For the latter we will consider the data from the 2dFGRS catalogue \cite{2dF}. For supernovae we will take two different compilations, the Constitution and SDSS samples, both calibrated with the MLCS2k2 fitter \cite{MLCS}. This choice is to avoid any bias resulting from the use of a fiducial $\Lambda$CDM model in the supernovae calibration, the case for instance of the Union2 compilation that is calibrated with the SALT2 fitter \cite{Union2}. Our joint analysis leads to the best-fit value $\alpha \approx -0.5$. This corresponds to a constant-rate dark matter creation, with the dark energy density decreasing linearly with the Hubble parameter \cite{Borges,tests,tests2,tests1,Zimdahl,tests3,tests4}. On the other hand, the value $\alpha = 0$, corresponding to the $\Lambda$CDM model, is comparatively rulled out, standing outside the 3$\sigma$ confidence interval. This might constitute, possibly for the first time, an evidence of particles creation at cosmological scale.

The paper is organised as follows. In the next section we discuss the dark degeneracy and the mapping of any general model of the dark sector into interacting two-component models. In section \ref{GCG} we particularise these results to the non-adiabatic GCG. In section \ref{Observations} we describe the performed observational tests, and the results are presented and discussed in section \ref{Results}, followed by conclusions in section \ref{Conclusions}.

\section{Dark degeneracy}
\label{DD}

Consider a dark fluid with equation of state $p = \omega \rho$, with $-1 \le \omega < 0$. Split this fluid into two components, one called dark matter, with zero pressure, and the other named dark energy, with equation of state $p_{\Lambda} = - \rho_{\Lambda}$. With help of the Friedmann equation\footnote{We are considering the spatially flat FLRW spacetime and doing $8\pi G = c = 1$.} $\rho=3H^2$, it is not difficult to show that
\begin{equation}\label{11}
\rho_{\Lambda} = -3\omega H^2.
\end{equation}
These components will in general interact. Let us define the rate of energy flux between them as
\begin{equation}\label{rate}
\Gamma = \frac{1}{na^3}\frac{d}{dt}(na^3),
\end{equation}
where $a$ is the scale factor and $n$ defines the dark matter particle number density. For non-relativistic matter, (\ref{rate}) can also be written as
\begin{equation}\label{26'}
\dot{\rho}_m + 3H \rho_m = \Gamma \rho_m,
\end{equation}
where $\rho_m$ is the dark matter density.
By using (\ref{11}), $\rho = 3H^2$ and the energy conservation equation
\begin{equation}\label{26}
\dot{\rho} + 3H (\rho + p) = 0,
\end{equation}
it is possible to show that the energy-flux rate is \cite{CB}
\begin{equation}\label{14}
\Gamma = \frac{\dot{\omega}}{\omega+1} - 3\omega H.
\end{equation}

The momentum transfer between the components is null in the isotropic background. If it is negligible at the perturbation level, we can prove that the dark energy component does not cluster \cite{CB,Maartens,Wands2}. This is true, in particular, if dark matter is non-relativistic. Therefore, the dark matter component can be identified with the observed clustering matter. This identification is necessary for observational tests involving the distribution of LSS, as we shall discuss in section \ref{Observations}. This result also guarantees that the dark matter power spectrum will not have oscillations or instabilities caused by pressure perturbations. Indeed, for any perfect fluid the evolution equation for the gravitational potential is given by \cite{GCG}
\begin{equation} \label{tam} 
\Phi_{B}''+3\mathcal{H}(1+c_{a}^2)\Phi_{B}' + \nonumber
[2\mathcal{H}'+(1+3c_{a}^2)\mathcal{H}^2+c_{s}^2k^2]\Phi_B = 0\,,
\end{equation}
where the prime means derivative with respect to $a$, $\Phi_{B}$ is the Bardeen gauge-invariant potential, $\mathcal{H}=aH$, $k$ is the perturbation comoving wavenumber, $c_a^2 = \partial p/\partial \rho$ defines the adiabatic sound velocity, and $c_s^2 = \delta p/\delta \rho$ is the effective sound velocity. If the dark energy component does not cluster, we have $\delta p = \delta p_{\Lambda} = 0$, leading to $c_s = 0$. Therefore, the term proportional to $k$ in (\ref{tam}) will be zero and the spectrum will be smooth. The difference between $c_a$ and $c_s$ means that the decomposed dark fluid is non-adiabatic.

\section{Non-adiabatic Chaplygin gas}
\label{GCG}

\begin{figure*}
\centerline{\includegraphics[height=6cm]{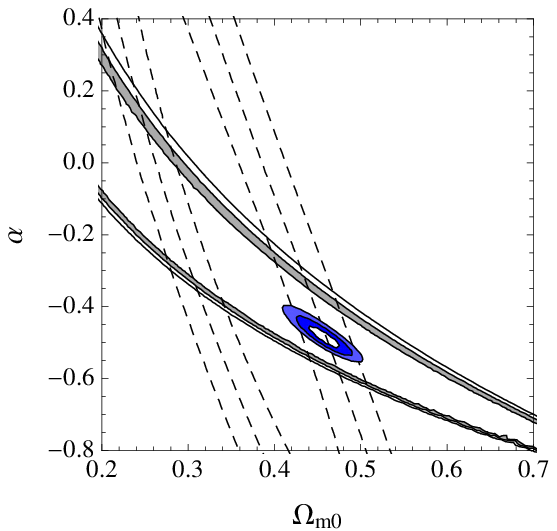} \hspace{.2in} \includegraphics[height=6cm]{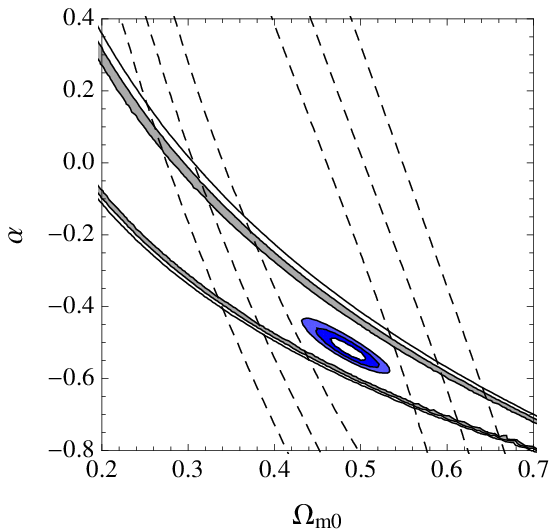}}
\caption{{\bf Left panel}: Confidence contours after marginalising over $H_0$. Dashed lines correspond to the Constitution sample of SNe Ia calibrated with MLCS2k2-17, solid lines correspond to the 2dFGRS power spectrum, and the blue ellipses correspond to the joint analysis of SNe Ia, the matter power spectrum and the position of the first peak of CMB. {\bf Right panel}: The same with the SDSS sample of SNe Ia calibrated with MLCS2k2.}
\label{fig1}
\end{figure*}

Using (\ref{EoS}) in the energy conservation equation (\ref{26})
we obtain
\begin{equation}\label{rho}
\rho = \rho_0 \left[ \bar{A} + \frac{1-\bar{A}}{a^{3(1+\alpha)}} \right]^{\frac{1}{1+\alpha}},
\end{equation}
where $\rho_0$ is the present GCG energy density and $\bar{A} \equiv A/\rho_0^{\alpha +1}$. This density scales with $a^{-3}$ for $a\ll 1$ and tends to a constant when $a\rightarrow \infty$. That is, at early times the gas behaves like dust matter, and tends to a cosmological constant in the asymptotic future. With $\alpha = 0$ we reobtain the $\Lambda$CDM model, with $\rho = \bar{A} + (1-\bar{A})/a^3$.
From (\ref{EoS}) we can derive the equation-of-state parameter
\begin{equation}\label{29}
\omega = \frac{p}{\rho} = -\frac{A}{\rho^{\alpha+1}},
\end{equation}
and hence the adiabatic sound speed
\begin{equation}\label{24}
c_a^2 = \frac{\dot{p}}{\dot{\rho}} = - \alpha \omega.
\end{equation}
By differentiating $p = \omega \rho$ we obtain
\begin{equation}\label{25}
\dot{\omega}\rho = -\omega (\alpha + 1) \dot{\rho}.
\end{equation}
Substituting into (\ref{26}) it follows that
\begin{equation}\label{27}
\dot{\omega} = 3 \omega (\alpha + 1) (\omega + 1) H.
\end{equation}

The gas can now be split into two interacting components like in the previous section, $\rho = \rho_m + \rho_{\Lambda}$, $p = p_{\Lambda} = -\rho_{\Lambda}$. Substituting (\ref{27}) into (\ref{14}) we obtain the rate of matter creation,
\begin{equation}\label{28}
\Gamma = 3 \alpha \omega H.
\end{equation}
Since $\rho = 3H^2$, we have, from (\ref{29}) and (\ref{28}),
\begin{equation}\label{30}
\Gamma = -\frac{\alpha A}{3^{\alpha}} H^{-(2\alpha+1)},
\end{equation}
while, from (\ref{11}) and (\ref{29}), it is easy to obtain
\begin{equation}\label{Lambda}
\rho_{\Lambda} = \rho_{\Lambda 0} \left( \frac{H}{H_{0}} \right)^{-2\alpha}.
\end{equation}
Since $\rho_{\Lambda} = A/\rho^{\alpha}$ (from (\ref{EoS})), we can see that $\Omega_{\Lambda 0} \equiv \rho_{\Lambda 0}/\rho_0 = \bar{A}$ and, therefore, $\Omega_{m0} = 1 - \bar{A}$. 
For $\alpha = 0$ we reobtain the $\Lambda$CDM model, with $\Gamma = 0$ and a constant $\rho_{\Lambda}$. When $\alpha < 0$ we have energy flux from dark energy to dark matter, since $\Gamma > 0$. For $\alpha = -1/2$ we have the particular case of a constant $\Gamma$, and $\rho_{\Lambda}$ decreases linearly with $H$. It was shown in \cite{GCG,Wands} that the split gas presents late-time non-adiabatic perturbations owing to the interaction between the components. This prevents the appearance of oscillations and instabilities present in the power spectrum of the adiabatic GCG \cite{Sandvik,Joras}, provided that the dark energy component is unperturbed \cite{GCG,Wands}. It was explicitly shown in \cite{Zimdahl} that this is true in the case $\alpha = -1/2$.

\section{Observational tests}
\label{Observations}

When testing the GCG against observations, we have to identify the clustering part of the gas to be compared to the observed amount of matter in galaxies and clusters. As discussed above, this identification is unique if we split the gas as in sections \ref{DD} and \ref{GCG}. We have seen that in this case the evolution equation (\ref{tam}) does not depend on $k$, which means that the profile of the dark matter power spectrum will depend essentially on the value of $k_{eq} \equiv H(z_{eq})/(1+z_{eq})$, the comoving horizon scale at the time of matter-radiation equality. By taking the limit $ a\ll 1$ in (\ref{rho}) we obtain, for the early-time matter density,
\begin{equation}\label{highz}
\rho_m = 3 H_0^2 \Omega_{m0}^{\frac{1}{1+\alpha}} z^3 \quad (z \gg 0).
\end{equation}
Comparing with the standard expression $\rho_m = 3 H_0^2 \Omega_{m0} z^3$ (case $\alpha = 0$) we see that, for the same amount of matter at high redshifts, we have more (less) matter today for a negative (positive) $\alpha$. This is expected because of the late-time matter creation (annihilation). With (\ref{highz}) and by using for radiation $\rho_{R} \approx 3H_0^2 \Omega_{R0} z^4$, we obtain the redshift of matter-radiation equality,
\begin{equation}
z_{eq} = \frac{\Omega_{m0}^{\frac{1}{1+\alpha}}}{\Omega_{R0}}.
\end{equation}
Hence, with the use of $3H^2 = \rho_m + \rho_R$,  we have
\begin{equation}
k_{eq} = \sqrt{\frac{2}{\Omega_{R0}}} \frac{\Omega_{m0}^{\frac{1}{1+\alpha}}}{l_{H0}} = 0.073 \text{Mpc}^{-1} h^2 \Omega_{m0}^{\frac{1}{1+\alpha}},
\end{equation}
where $l_{H0} = 3000 h^{-1}$ Mpc is the present Hubble radius, and we are fixing the present radiation density parameter at the standard value $\Omega_{R0} = 4.15 \times 10^{-5} h^{-2}$. For $\alpha = 0$ the above expressions are reduced to the standard model ones (see, for instance, equation (7.39) of Ref. \cite{Dodelson}).
The present power spectrum is, apart from a normalisation constant, given by ${\cal P} =  k T^2(k)$, if we take the scalar spectral index $n_s = 1$. Neglecting baryons (whose density is only about $5\%$ of the total density) the BBKS transfer function $T(k)$ is given by \cite{BBKS}
\begin{widetext}
\begin{equation}\label{transfer}
T(x = k/k_{eq}) = \frac{ln[1+0.171x]}{(0.171x)}  \left[1 + 0.284x + (1.18x)^2  + (0.399x)^3 + (0.490x)^4\right]^{-0.25}.
\end{equation}
\end{widetext}
This transfer function was obtained by precisely fitting the observed spectrum with the CDM model. It gives the spectrum profile at the begining of the matter dominated era and, therefore, is valid for any model that differs from the CDM model only at late-times\footnote{It is instructive to compute the Hubble parameter and the matter density at the redshift of matter-radiation equality in the standard model ($\alpha = 0$) and in a case with matter creation (say, $\alpha = -1/2$). Taking in the first case $\Omega_{m0} \approx 0.2$ (the best-fit given by LSS observations) and in the second case the best-fit value $\Omega_{m0} \approx 0.45$, the redshift of equality is $z_{eq} \approx 2,400$ in both cases. At this redshift the matter densities differ by $1\%$, and the Hubble parameter in the two models differ by $0.3\%$.}. During the late-time expansion the spectrum is just amplified, and for its normalisation we use the observed spectrum at small scales.

The observed position of the first acoustic peak in the CMB anisotropy spectrum is given by \cite{Hu} $l_1 = l_A (1-\delta_1$), where $l_A$ is the acoustic scale and 
\begin{equation}
\delta_1 \approx 0.267 \left( \frac{r}{0.3} \right)^{0.1}.
\end{equation}
Here, $r \equiv \rho_R(z_{ls})/\rho_m(z_{ls})$ is evaluated at the redshift of last scattering $z_{ls}$\footnote{We have shown in \cite{Cassio} that, for the particular case $\alpha = -1/2$, $z_{ls}$ differs only $1\%$ from the standard value.}. Using again (\ref{highz}) and $\rho_R \approx 3H_0^2 \Omega_{R0} z^4$, we have
\begin{equation}
r = \Omega_{R0} \Omega_{m0}^{-\frac{1}{1+\alpha}} z_{ls}.
\end{equation}
For calculating the acoustic scale, we use $3H^2 = \rho + \rho_R$. The density parameters of baryons and photons, to be used in the calculation of the sound speed in the baryon-photon plasma, will be fixed at the same values of the standard model, $\Omega_{b0} h^2 = 0.02$ and $\Omega_{\gamma 0} h^2 = 2.45 \times 10^{-5}$.

Finally, when testing the distance-redshift relation of type Ia supernovae, the calibration method has special relevance. The Union2 sample was calibrated with the SALT2 fitter, and for it a fiducial $\Lambda$CDM model was used \cite{Union2}, something that may lead to bias. A model-independent calibration is provided by the MLCS2k2 fitter, used in the calibration of the Constitution and SDSS compilations to be considered here \cite{MLCS}.

\section{Results and discussion}
\label{Results}

\begin{figure}
\centerline{\includegraphics[height=6cm]{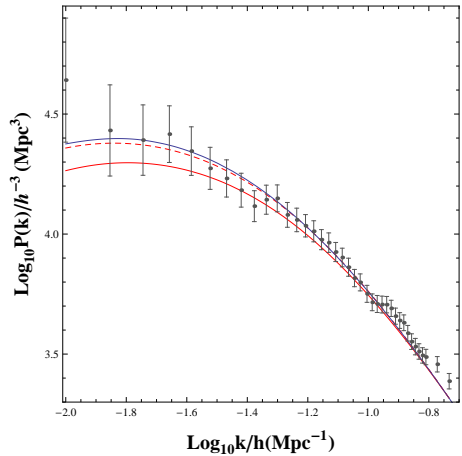} \hspace{.2in}}
\caption{Power spectrum for $\alpha = -1/2$ and $\Omega_{m0} = 0.45$. The blue line was obtained by integrating the perturbation equations \cite{tests2}. The dashed red line corresponds to the use of the transfer function (\ref{transfer}). For comparison, we also show the power spectrum of the $\Lambda$CDM concordance model (solid red line). Data are from the 2dFGRS catalogue.}
\label{fig2}
\end{figure}

\begin{table}
\begin{center}
\caption{$3\sigma$ intervals for $\alpha$ and $\Omega_{m0}$ (SNe Ia+CMB+LSS).} 

\begin{tabular}{rcccc}
\hline \hline \\
\multicolumn{1}{c}{SNe Ia sample}&
\multicolumn{1}{c}{$\alpha$}&
\multicolumn{1}{c}{$\Omega_{m0}$}&
\multicolumn{1}{c}{$\chi^2_{min}/\nu$}\\ \hline \\
Constitution (MLCS2k2-17) &$-0.49^{+0.09}_{-0.05}$ & $0.46^{+0.04}_{-0.05}$ & $0.98$ & \\
SDSS (MLCS2k2) & $-0.51^{+0.08}_{-0.07}$ & $0.48\pm 0.05$ & $0.85$ &  \\
SDSS (SALT2) & $-0.36^{+0.16}_{-0.08}$ & $0.36^{+0.05}_{-0.08}$ & $0.88$ & \\
Union2.1 (SALT2) & $-0.36^{+0.16}_{-0.11} $ & $0.36^{+0.06}_{-0.09}$ & $0.94$ & \\
\hline \hline
\end{tabular}
\end{center}
\end{table}

Table I shows the joint analysis results after marginalising over $H_0$, with $3\sigma$ confidence level. The two first lines correspond to the SNe Ia samples calibrated with MLCS2k2, and in both cases we obtain $\alpha \approx -1/2$, which corresponds to a constant-rate production of dark matter and to a dark energy density decaying linearly with $H$. The corresponding confidence regions are show in Fig. \ref{fig1}. The present matter density is in agreement with concordance tests with $\alpha$ fixed as $-1/2$, which have given $\Omega_{m0} \approx 0.45$ \cite{tests}. It is higher than the standard model best-fit, as antecipated above for the case of a negative $\alpha$. The power spectrum for $\alpha = -1/2$ and $\Omega_{m0} = 0.45$ is shown in Fig. \ref{fig2}, together with the 2dFGRS data. The blue line was obtained by integrating the perturbation equations from very high redshifts to the present, as in Ref. \cite{tests2}. The dashed red line corresponds to the use of Eq. (\ref{transfer}) and leads to an error of $4\%$ at most. For comparison we show the spectrum of the $\Lambda$CDM concordance model (solid red line). The GCG best fit corresponds to a Universe age $t_0 \approx 13.5$ Gyr (for $h \approx 0.7$) \cite{tests}.

The value $\alpha = 0$, which corresponds to the $\Lambda$CDM model, is rulled out in comparison with the best fit, standing outside the $3\sigma$ confidence interval (see, however, Ref. \cite{Wands,WandsGCG}). This agrees with previous tests in which $\alpha$ was fixed as zero and that gave high values of $\chi^2_{min}$ for the SDSS and Constitution samples \cite{tests}. This relatively poor concordance reflects the fact that the matter density obtained with LSS alone is $\Omega_{m0} \approx 0.2$ \cite{2dF,SDSS}, while the best fit obtained with SNe Ia tests using the Constitution and SDSS samples calibrated with MLCS2k2 gives $\Omega_{m0} \approx 0.3$ and $\Omega_{m0} \approx 0.4$, respectively (see Table II of Ref. \cite{tests1}). As we have discussed in the previous section, LSS tests strongly depend on the matter density at high redshifts, while SNe Ia probes are sensitive to the matter density at low redshifts. Therefore, this tension between the LSS and SNe Ia values of $\Omega_{m0}$ may suggest a late-time process of matter creation.

\begin{figure}
\centerline{\includegraphics[height=6cm]{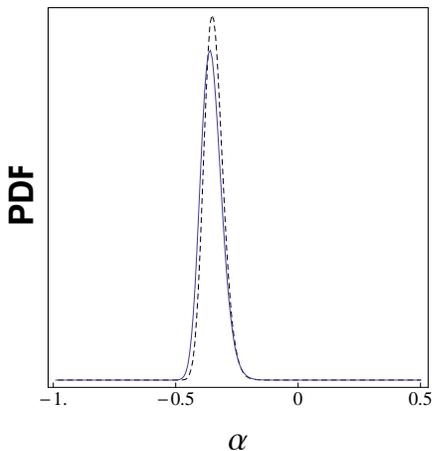} \hspace{.2in}}
\caption{Probability distribution functions for $\alpha$ for the joint analysis of SNe Ia + LSS + CMB. The solid (dashed) line corresponds to Union2.1 (SDSS) supernovae sample calibrated with SALT2.}
\label{fig3}
\end{figure}

In the last two lines of Table I we show, for comparison, the joint analysis results when we use the SDSS and Union2.1 samples of SNe Ia calibrated with SALT2. As antecipated in the previous section, the results are moved towards $\alpha = 0$. However, the $\Lambda$CDM model ($\alpha = 0$) is still outside the $3\sigma$ confidence intervals. In Fig. \ref{fig3} we show the corresponding probability distribution functions after marginalising over $h$ and $\Omega_{m0}$.

To finish this analysis we need to comment about the power spectrum normalisation. Since the dark energy component does not cluster, the dark matter production is homogeneous, and this causes a late-time suppression of the dark matter density contrast $\delta_m \equiv \delta \rho_m/\rho_m$ (see the left panel of Fig. 3 in Ref. \cite{GCG}). So, why is the spectrum normalisation not affected? To answer this question, let us remark that we observe the baryonic power spectrum, not the dark matter one. Since baryons are conserved, no suppression in their spectrum is expected, although it presents the same profile as the dark matter spectrum, determined by $k_{eq}$. This can also be understood if we remember that the baryons distribution follows the gravitational potential, which is determined (via Poisson's equation) by the dark matter perturbations $\delta \rho_m$, not by the contrast $\delta_m$ itself. The reader can see in the right panel of Fig. 3 in Ref. \cite{GCG} that the gravitational potential for $\alpha = -1/2$ presents approximately the same strength and evolution as in the $\Lambda$CDM model. This guarantees, by the way, that the CMB power spectrum will also have the correct normalisation. An analysis of the $\alpha = -1/2$ case with the inclusion of conserved baryons can be found in \cite{Rodrigo}. A study of non-linear collapse in this model was recently done in \cite{tests4}.

\hspace{6in}

\section{Conclusions}
\label{Conclusions}

Models with interaction in the dark sector have been proposed for a long time (see \cite{Chimento,Chimento2} and references therein). In Ref. \cite{CB} it was shown that, in fact, any dark sector model can be mapped into a non-adiabatic fluid with energy flux between the dark energy and dark matter components. Choosing, on the basis of observations, a specific interaction inside this class of models is a difficult task, since the rate $\Gamma$ is a general function, related to the function $\omega$ by Eq. (\ref{14}). This task becames simpler if we particularise the dark sector to a non-adiabatic GCG, since now the rate $\Gamma$, given by (\ref{30}), is known up to the constant parameter $\alpha$, which can be determined from observations. Our analysis has led to a constant rate of matter creation, which corresponds to a dark energy density that decays linearly with the Hubble parameter. Interestingly enough, a linear dependence on $H$ is expected from estimations of the energy density of the QCD vacuum condensate in the FLRW spacetime \cite{Schutz}. On the other hand, from (\ref{28}) it is easy to see that the obtained constant rate is equal to $3H_{dS}/2$, where $H_{dS}$ is the Hubble parameter in the de Sitter limit. In other words, the creation rate is equal, apart from a factor of the order of unity, to the Gibbons-Hawking temperature of the future de Sitter horizon. We have also seen that the $\Lambda$CDM model is comparatively rulled out, an evidence in favor of matter creation at cosmological scale. This evidence may be confirmed with future observations, in which the discrepancies between the standard and interacting models will be more evident. For example, in Ref. \cite{tests4} a slightly higher number density of high mass structures is predicted for $\alpha = -1/2$ as compared to the standard case. Although inside the error bars of present observations, this difference may be tested in future surveys like J-PAS \cite{j-pas} and eROSITA \cite{eRosita}.

${ }$


We are thankful to J. C. Fabris, H. Velten, I. Waga, D. Wands and J. Gonz\'alez for discussions and help. This work was partially supported by CNPq (Brazil).

\end{document}